# DYNAMICS OF THE VORTEX FLUID IN CUPRATE SUPERCONDUCTORS

## ABSTRACT

We calculate the Nernst effect in the vortex fluid phase, which occurs in the lower-T portion of the pseudogap region of the high Tc cuprate phase diagram. The dynamics is dominated by the flows due to both thermally excited vortices and those caused by the magnetic field; we show that the flow of the latter due to the thermal gradient encounters a viscous force caused by the random vortex flows. The temperature and field dependence is controlled by the effects of T and B on the momentum-dependent energy gaps and thence on the superfluid density. A reasonable level of agreement with the observational data is obtained, in view of the crudity of the assumptions.

______________________________________________________________

It now seems clear that the region of the phase diagram of cuprate superconductors where vorticity effects occur, first glimpsed by Orenstein[1], and interpreted as a vortex fluid and explored in depth by Ong[2], is a phase in which the d-wave gap persists but loses phase coherence. This is a consequence of the spin-charge "locking" mechanism,[3] in that the superconducting gap is established as a fixed structure in the spin space but only the "extended s" part of the gap is locked to the "charge" axis of the real electrons, while the phase of the real electrons fluctuates in space and time. (This is more specific than the "QED III" idea which leaves the microscopic mechanism indeterminate).

A phase in which the d-wave gap persists has variable gapping around the Fermi surface and therefore some low-energy degrees of freedom which are quasiparticles. Therefore $\rho_s$ and $H_{c2}$ vary around the Fermi surface and with temperature. Presumably there are also particle currents but the Homes rule physics [ref 3] suggests that these are not very important near Tc. We will assume that the important currents are supercurrents, ignoring electrodynamical effects of quasiparticles since these can hardly exhibit the pronounced nonlinear responses which are observed. In particular, the random motions of vortices are caused by the supercurrents of other vortices.

The Bose fluid which is the correct model for this state is often seen as equivalent to the x-y model but is not quite, although its critical behavior in the absence of a field is x-y because that depends on the longest wavelength correlations. To understand the dynamics it is essential to take into account the incompressibility of the electron fluid; the order parameter is not allowed to fluctuate arbitrarily but must satisfy charge conservation:

$$\nabla \cdot j = 0; \text{ therefore, since } j \propto \nabla \varphi,$$
$$\nabla^2 \varphi = 0 \qquad [1]$$

φ being the fluctuating phase of the d-wave gap. Therefore the state is a vortex fluid: it can be completely characterized by a fluctuating array of vortices, singular lines of the phase (or in 2D, points). The reason for charge conservation is that the material is a metal and therefore charge fluctuations can only persist for times of the order of the inverse plasma frequency, while the vortex motions are at frequencies of the order kT/h.

For the time being we confine our discussion to 2D which amounts to neglecting the transverse stiffness of the vortices. Most samples are approximately 2D. I think doping makes more difference to the phenomena than dimensionality, but I suspect dimensionality is important at small B, since at large length scales coherence in the third dimension takes over.

Previous work[4] has treated the vortex fluid as a problem in fluctuations near a critical point, and these papers are valid near the critical temperature. But the concept here is to treat the vortex fluid phase as a thermodynamic phase in itself with characteristic properties specific to it as a fixed point, distinct from those of a normal metal. (This point of view has been anticipated by Sudbo[5] and Feigel'man[6]). I choose to calculate the vortex Nernst effect which on examination seems to be simpler, because it is small near the critical temperature and therefore critical behavior isn't important; also, it runs smoothly into the high field, low T region where we understand it quite well. In the Nernst effect, we can think of the vortices being driven by the temperature gradient and encountering a viscous damping which, like the resistivity, is caused by the fluctuating supercurrents—in fact, it is well-known that the viscous damping coefficient is proportional to the conductivity. Without damping, the Nernst effect vanishes.

There are two additional considerations which give this point of view a different slant from ref 4. First, the vital energy is not that of interaction of the vortices but their self-energy; second, there is a Gauss's theorem—in the equivalent charge gas model—or a phase theorem in the vortex model—which has the result that there is no effective screening of the currents due to the excess vortices caused by the B-field.  These excess vortices do not experience Kosterlitz-Thouless screening and retain their entire 1/r current structure until one reaches the magnetic length, at which they are compensated by the vector potential of the B-field.  This fact requires either a lot of cogitation, or Ragu's simulations[7], to convince oneself of, but is true.  The supercurrent is proportional to $\nabla\phi$ and the excess phase gradient and excess current caused by the single discrete vortex for each flux quantum cannot be made to go away at any radius.  That is, imagining that one has a region of area $l_B^2$ which contains n+1 "+" vortices and n "-" ones, try as one may to pair them off in such a way as to screen away the extra currents, it is impossible: the extra vortex remains quantized and cannot be distributed over the entire region.  But it's important to realize that it is not possible to identify an "extra" vortex with any particular one of the thermal gas of vortices which are present: it does not cause any inhomogeneity of the vortex density greater than the fluctuations which already exist, even though its contribution to the energy diverges logarithmically with $l_B$.  In this sense this phase can be characterized unequivocally, although clearly it has no order parameter in the conventional sense.

From reference 1 we borrow the equation
$$e \equiv E/\nabla T = |B \times v|/\nabla T = BS_\phi/\eta \quad [2]$$
where e is the Nernst coefficient defined as in the equation, $S_\phi$ is the "entropy" per vortex, and $\eta$ is the viscous damping coefficient for a vortex as it moves through the medium.. Equation [2] comes from balancing the viscous force $\eta v$ against the thermal force $S_\phi \nabla T$. The damping coefficient we estimate in the following way.  First, we estimate the diffusion constant for a random vortex fluid.  A typical point in the fluid (including a vortex core) will be moving at a velocity v=h/ml, where l is half the typical distance between vortices ($l^{-2}$ is, neglecting order 1 factors, the density of thermal vortices $n_V$ ).  It will continue to move in the same direction for a time $\tau$=l/v. Its diffusion constant is thus

$$D = l^2/\tau = vl = h/m \quad [3]$$

(This is actually a familiar result.) We use the Einstein argument to relate D to η: In the presence of a gradient in chemical potential of the vortices, which is a force field, there will be a gradient in concentration such that the diffusion current cancels the flow due to the force:

$$D \frac{dn_V}{dx} = \eta^{-1} \frac{d\mu_V}{dx} ; \therefore \eta = D^{-1} \frac{d\mu_V}{dn_V} = \frac{mkT}{h} \quad [4]$$

From the argument that the thermal energy of the pair currents must be ~ kT, which is checked, incidentally, by the Homes relation kTc=h/τ, we may substitute, not too far from Tc,

$$kT = \hbar^2 / ml^2, \text{ so that } \eta = \hbar n_V \quad [5]$$

Finally, using equation [2], we have that

$$e = \frac{B}{\phi_0} \frac{\phi_0 S_\phi}{\hbar n_V} = \frac{n_B}{n_V} \frac{c}{e} S_\phi \quad [6]$$

(the e in c/e is the electron charge, and the factor disappears when we quote the Nernst coefficient e in volts.) $n_B$ is the number of added vortices due to B. Boltzmann's constant is about 120 μV/degree, and at the maximum of the Nernst effect near Tc, (as we shall shortly demonstrate) $S_\phi$ is an average of logarithms which turns out to be of order 1/3 or less, so that the crucial factor is the ratio of vortex densities. This is about an order of magnitude: At 100 degrees, the thermal $n_V$ is about 1.5 $10^{13}$ while at 15T $n_B$ is 1/10 of that. Maximum e's in typical cases from ref 2 are around 3-5 μV/degree, and this is indeed about 1/30 $k_B$. If anything, our estimates are a little high, and this is appropriate because as we shall see the supercurrents due to considerable portions of the Fermi surface may already be reduced by quasiparticle excitations, especially near optimal doping.

The excess energy of the vortices caused by the B-field is

$$U_V = (B/\Phi_0)\rho_s(\Delta,T)\ln\frac{H_{c2}}{B}; \quad B < H_{c2}$$

$$U_V = 0, \quad B > H_{c2} \quad [7]$$

for each patch of the Fermi surface with gap value Δ.
Near Tc the energy and entropy terms in the free energy almost exactly cancel, since the transition is due to the proliferation of thermal vortices. Thus it doesn't much matter whether we consider the driving force of the Nernst effect to be entropy or energy, but in fact the thermodynamic identity of the Nernst and Ettingshausen coefficients tells us that $S_\phi$ is really the energy [7] divided by T. From this point on we do not keep explicit track of the dimensional multiplying factors and concentrate instead on the variation of this energy with T and B.

We will label the different Fermi surface patches with their gap values and think of [2] as to be summed over the different patches with a distribution p(Δ).

For a simple model we can use the sinusoidal distribution

$$p_0(\Delta)d\Delta = 1/\sqrt{(1-\Delta^2/\Delta^2_{max})}d\Delta/\Delta_{max} \quad [8]$$

but to model the tunneling data it may be necessary to add in a δ-function at $\Delta_{max}$. Δ itself, we shall assume, is not much dependent on B, an assumption which may fail for overdoping where we are near to T*.

$H_{c2}$ is of course proportional to $\Delta^2$:

$$H_{c2} = \Phi_0/\xi^2 = (c/2e\hbar v_F^2)\Delta^2 = K\Delta^2,$$
(defining K) [9]

And $\Delta_{max}$ itself may be a fairly strong function of T, especially for optimal to overdoping.

$\rho_s$ 's dependence on $\Delta/T$ is calculated from the expression

$$\rho_s/\rho_s(T=0) = 1 - \int d\varepsilon_k \frac{df(E_k)}{dE_k} \quad [10]$$

where f is the Fermi function. It is easy to establish that (10) varies as $(\Delta/T)^2$ for small values of $\Delta$ and as $1-\exp-(\Delta/T)$ for large. This means that small values of $H_{c2}$ are suppressed increasingly with T, which I suspect will make the logarithmic singularity hard to see experimentally for T>Tc. In fact, [10] is an underestimate because the time-variation of the phase of $\Delta$ leads to a motional narrowing of its effect on the quasiparticles, and effect which is responsible for the "Fermi arc" observations of Campuzano[8] and others. Fortunately, this effect occurs also at $\Delta \approx 1/\tau \approx kT$, so that this merely enhances the steepness of the dependence on $\Delta/T$.

Thus the final expression for the Nernst energy, the energy which is delivered per square cm by the moving vortex structure, is

$$\nu \propto B \int_{\sqrt{B/K}}^{\Delta_{max}(T)} p_0(\Delta) d\Delta \, \rho_s(\Delta/T) \ln(K\Delta^2/B) \quad [11]$$

In view of the steep, Fermi-like dependence of $\rho_s$ on $\Delta/T$, and the many approximations so far, it is OK to replace the $\rho_s$ dependence with a second lower cutoff of the integral, $\Delta > \max[T, \sqrt{B/K}]$. In fig 1 I present the result of a very crude model calculation of [11] using a constant $p_0(\Delta)$ and using the step-function approximation to $\rho_S(\Delta/T)$.

All the formulas above are more accurate in cases such as one-layer BSSCO which are strongly two-dimensional, so that the sharp, Kosterlitz-Thouless transition occurs. Critical behavior near Tc, and vortex tortuosity for small B, have been neglected and may modify the behavior especially near Tc, so that the results of these formulas can not be expected to be exact numerical predictions, but as far as I can see they do indicate the right magnitude and general trend of the effect.

I would like to acknowledge extensive discussions of the data with N P Ong and Yayu Wang, and permission to use fig 2 from the latter. I acknowledge discussions of the theory with Sri Ragu, D Huse, Vadim Oganesyan and V Muthukumar, and assistance with fig 1 from B Edegger

FIGURE CAPTIONS

Figure 1  Theoretical Nernst coefficient at increasing temperatures starting with Tc from equation [11] neglecting T-dependence of $\rho_s$. B is in units of $H_{c2}$(max), v (the abscissa) in arbitrary units.

Figure 2  Nernst coefficient as a function of temperature from ref for some two-dimensional samples with lower Tc's. A few data below Tc is shown to emphasize continuity with conventional vortex liquid state. Note resemblance to figure 1 in steepness at B=0, convergence of curves at high B, and of course in general shape; detailed fits would be premature.

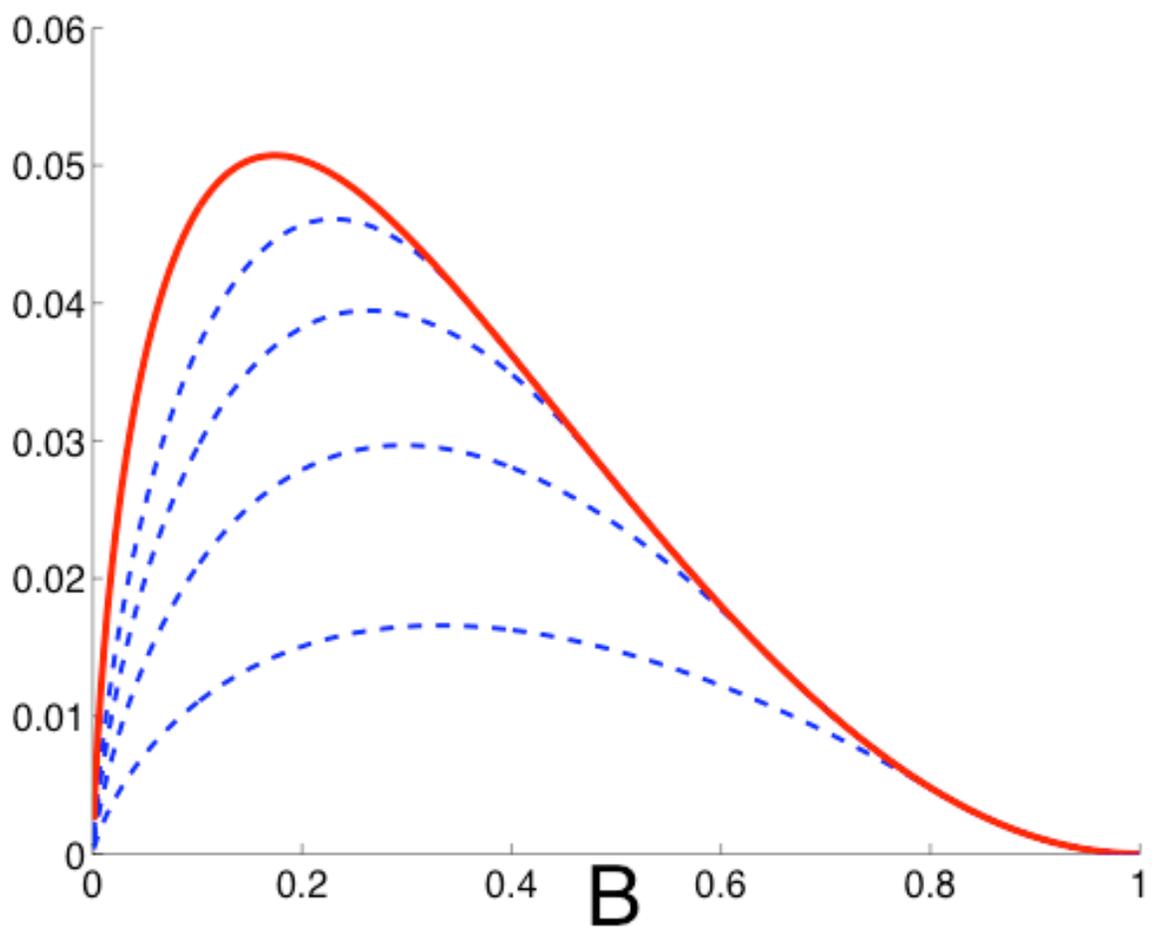
figure 1

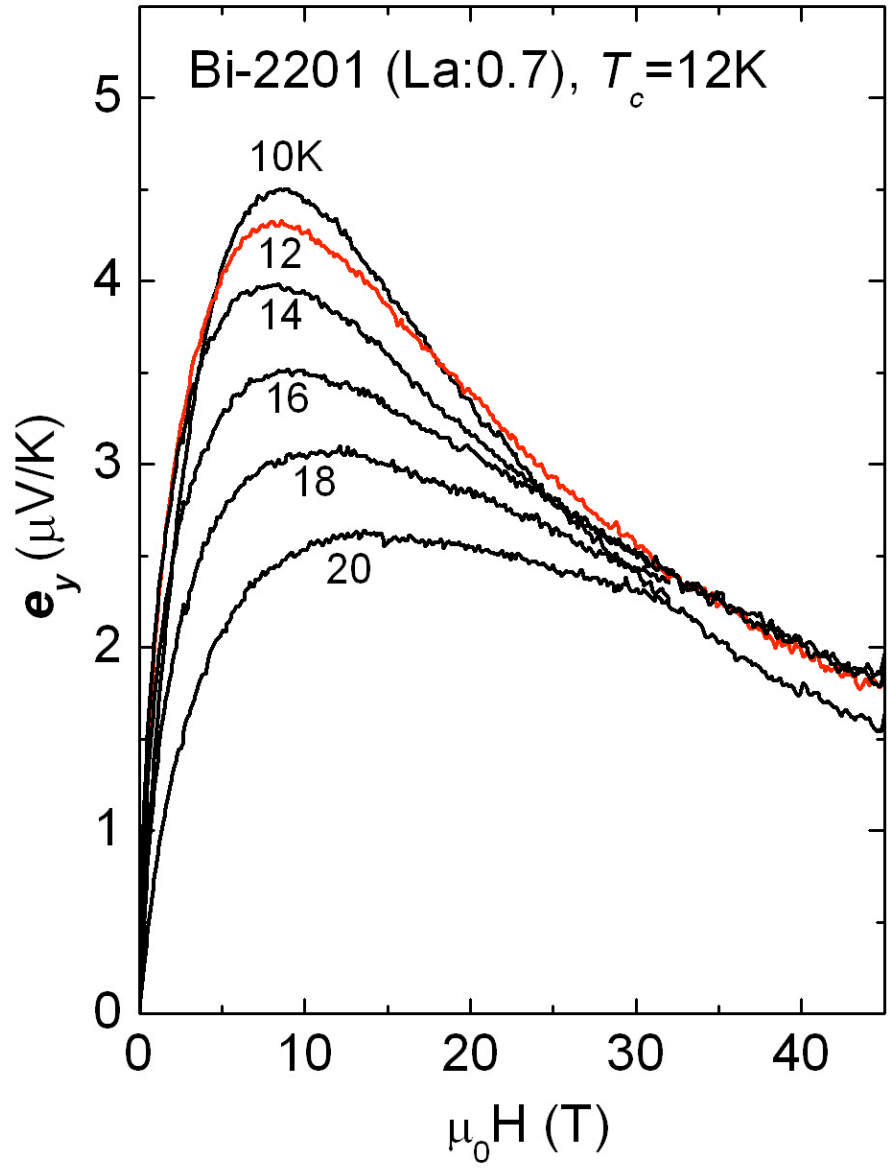

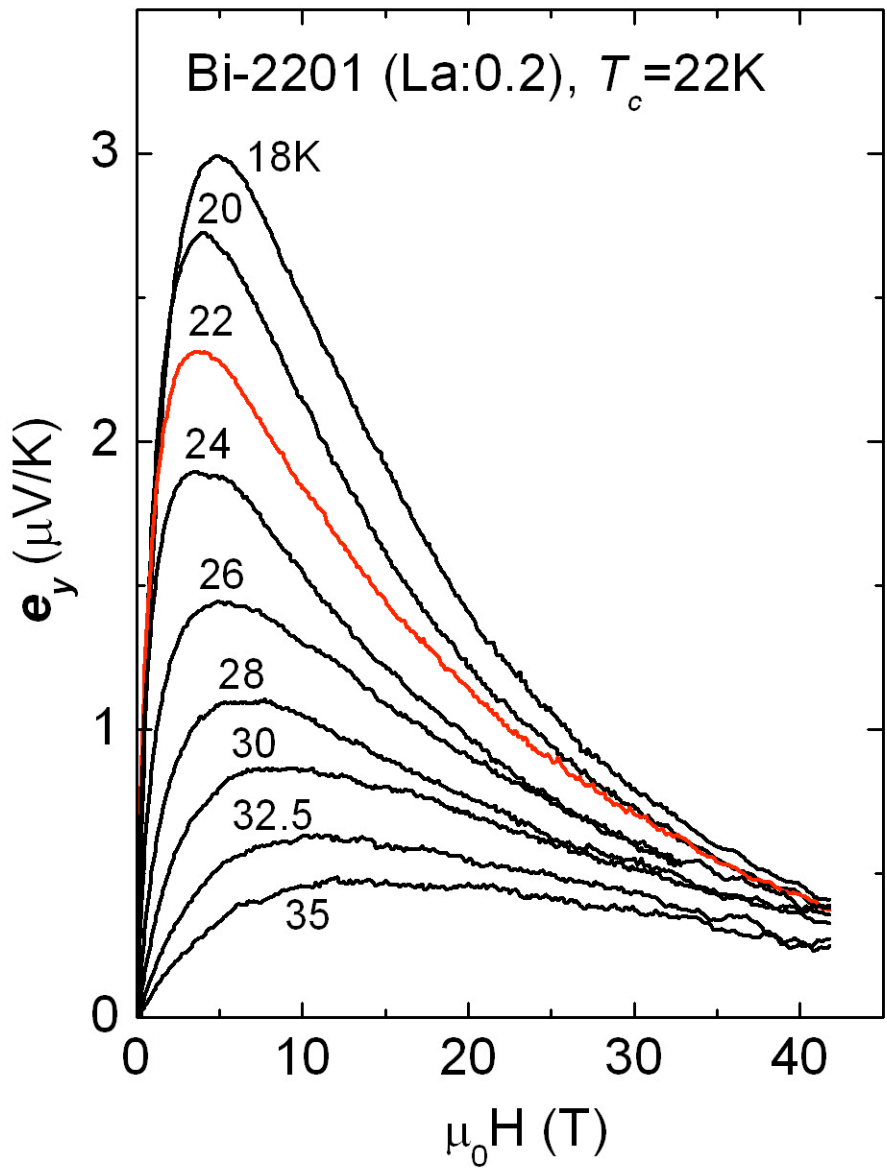

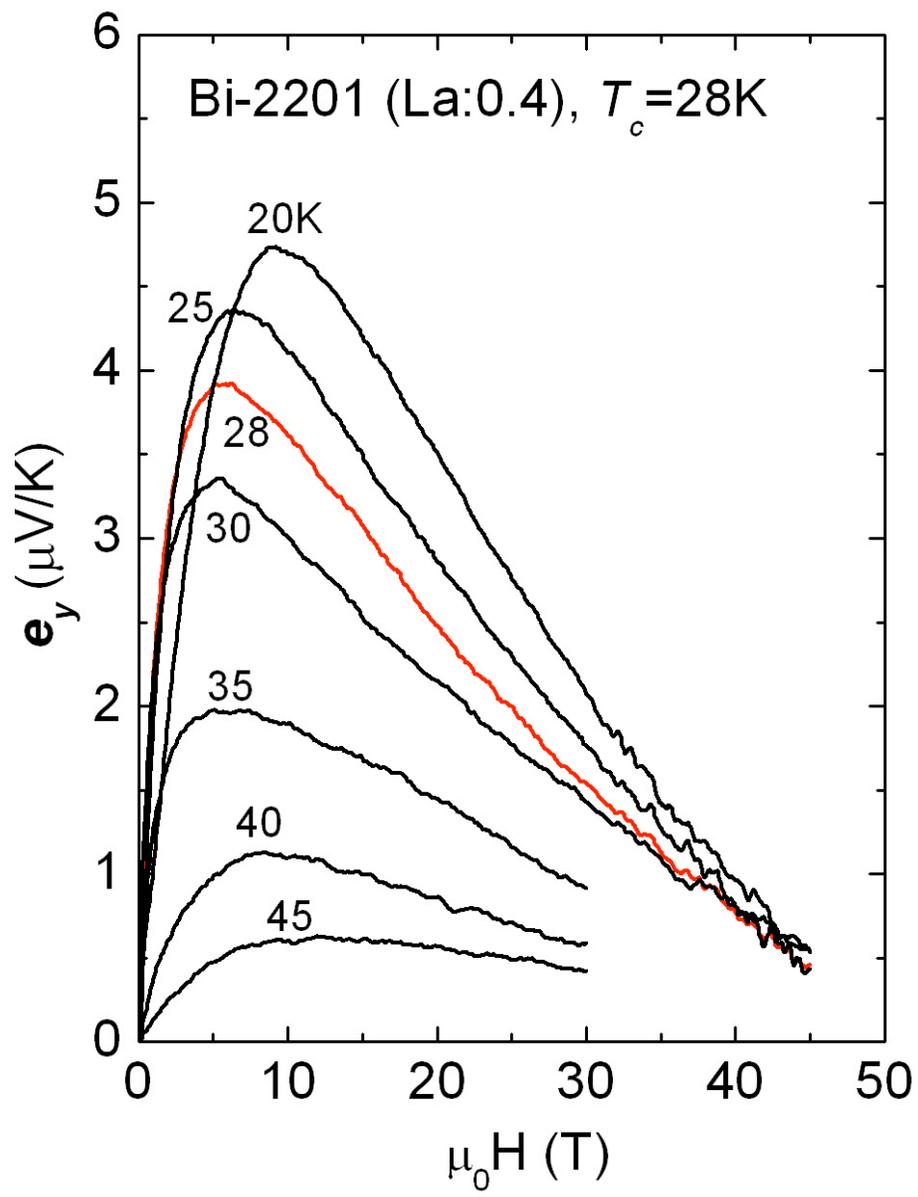

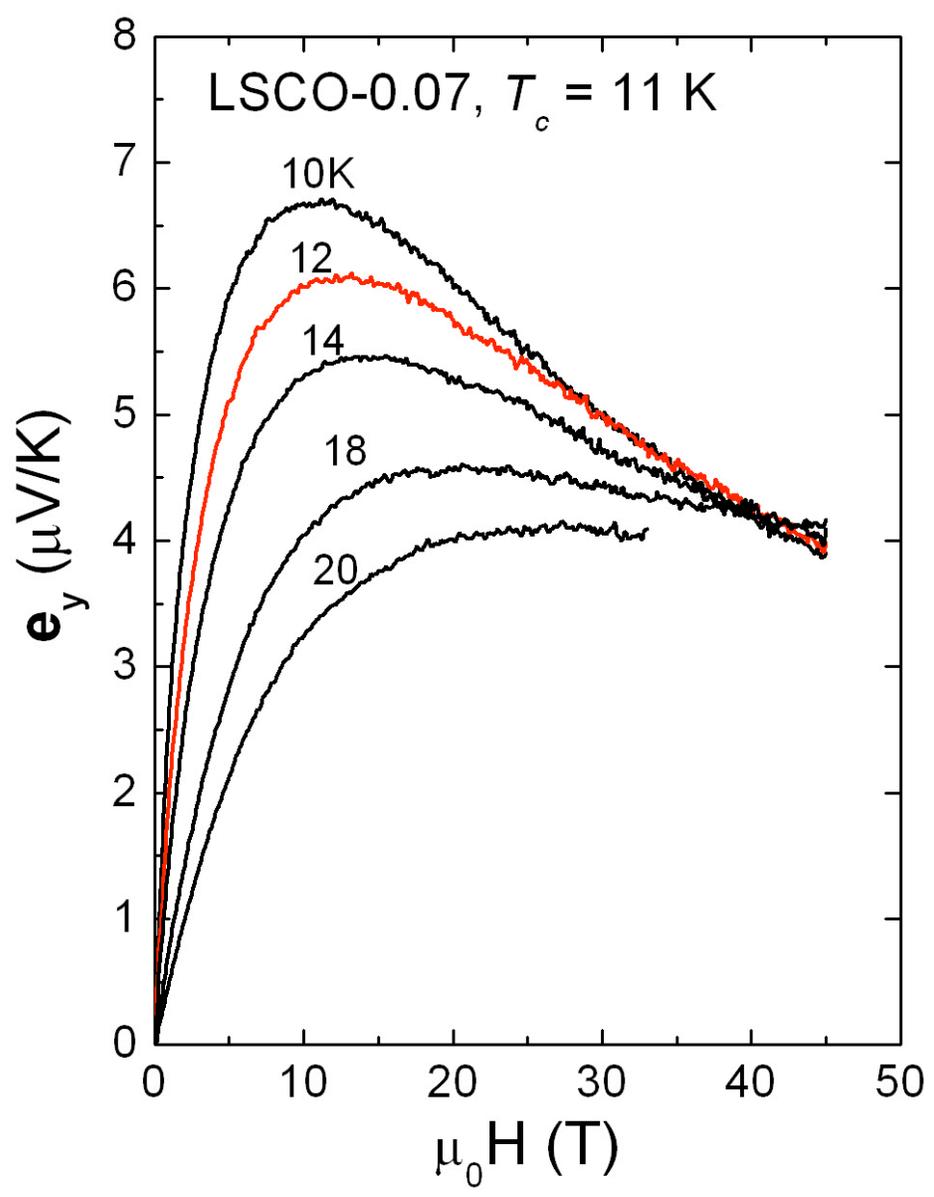